\documentclass[twocolumn, secnumarabic, amssymb, nobibnotes, prl, superscriptaddress]{revtex4}

\usepackage{graphicx} 
\usepackage{amsmath}
\usepackage{float}
\usepackage{color}
\usepackage[english]{babel}
\usepackage{array}
\usepackage[T1]{fontenc}
\usepackage[usenames,dvipsnames]{xcolor}
\usepackage{setspace}
\usepackage{hyperref}
\usepackage{bm}
\usepackage{lipsum} 
\usepackage{filecontents}
\usepackage{tikz}
\usepackage{stmaryrd}
\makeatletter
\renewcommand*{\fnum@figure}{{\normalfont \small{FIG.}~\thefigure}}
\makeatother
\begin{document}


\title{Logarithmic aging via instability cascades in disordered systems}
\author{Dor Shohat}
\email{dorshohat@mail.tau.ac.il}

\affiliation{Department of Condensed Matter, School of Physics and Astronomy, Tel Aviv University, Tel Aviv 69978, Israel}
\affiliation{Center for Physics and Chemistry of Living Systems, Tel Aviv University, Tel Aviv 69978, Israel}

\author{Yaniv Friedman}
\affiliation{Department of Condensed Matter, School of Physics and Astronomy, Tel Aviv University, Tel Aviv 69978, Israel}
\affiliation{Center for Physics and Chemistry of Living Systems, Tel Aviv University, Tel Aviv 69978, Israel}

\author{Yoav Lahini}
\email{lahini@tauex.tau.ac.il}
\affiliation{Department of Condensed Matter, School of Physics and Astronomy, Tel Aviv University, Tel Aviv 69978, Israel}
\affiliation{Center for Physics and Chemistry of Living Systems, Tel Aviv University, Tel Aviv 69978, Israel}

\begin{abstract}
Many complex and disordered systems fail to reach equilibrium after they have been quenched or perturbed. Instead, they sluggishly relax toward equilibrium at an ever-slowing, history-dependent rate, a process termed physical aging. The microscopic processes underlying the dynamic slow-down during aging and the reason for its similar occurrence in different systems remain poorly understood. Here, we reveal the structural mechanism underlying logarithmic aging in disordered mechanical systems, through experiments in crumpled sheets and simulations of a disordered network of bi-stable elastic elements. We show that under load, the system self-organizes to a metastable state poised on the verge of an instability, where it can remain for long, but finite times. The system’s relaxation is intermittent, advancing via rapid sequences of instabilities, grouped into self-similar, aging avalanches. Crucially, the quiescent dwell times between avalanches grow in proportion to the system's age, due to a slow increase of the lowest effective energy barrier, which leads to logarithmic aging.

\end{abstract}
\maketitle
\textbf{Introduction - } Understanding the unusual, seemingly universal dynamics of disordered systems trapped far from equilibrium remains a major challenge in condensed matter physics. Yet, they are not very hard to observe. Take, for example, a thin plastic sheet and crumple it into a ball. It might not be immediately apparent, but this seemingly mundane object exhibits many of the hallmark behaviors shared by non-equilibrium complex and disordered systems. These include slow relaxations and aging \cite{matan2002crumpling, lahini2017nonmonotonic}, intermittent mechanical responses \cite{kramer1996universal, shohat2022memory} emission of broadly-distributed crackling noise \cite{kramer1996universal,houle1996acoustic}, and a range of memory effects \cite{keim2019memory, matan2002crumpling,lahini2017nonmonotonic,Oppenheimer2015shapeable,shohat2022memory,shohat2023dissipation}. 

Perhaps the most remarkable behaviour is the physical aging of crumpled sheets. If placed under constant external load, a crumpled ball seems to never reach equilibrium. Instead, it compacts in an ever-slowing, logarithmic manner spanning many time scales, from fractions of a second to several weeks \cite{matan2002crumpling}. Here, the instantaneous relaxation timescales, indicated by both the compaction rate and the rate of emitted crackling noise, grow in proportion to the system's age -  the time passed since the system has been placed under the external load \cite{lahini2022Crackling}.

Similar physical aging dynamics is exhibited by a strikingly wide range of complex and disordered systems \cite{2003SRaN}. Examples span disordered conductors and superconductors \cite{gurevich1993time, vaknin2000aging, grenet2007anomalous}, biological systems \cite{morgan2020glassy, lieleg2011slow, kaplan2021observation}, disordered interfaces \cite{ben2010slip,kaz2012physical}, and soft amorphous materials \cite{song2022gels,knight1995density}. Diverging relaxation times are also a hallmark of glass-forming systems \cite{angell1995formation,debenedetti2001supercooled}, which exhibit slow logarithmic aging below their glass transition temperature \cite{Arceri2020,lundgren1983dynamics,struik1977physical,qiao2014dynamic,weeks2000three,samarakoon2016aging}. Yet, despite substantial theoretical and phenomenological studies \cite{2003SRaN, narayanaswamy1971model,bouchaud1992weak,cugliandolo1993analytical,sibani1989hierarchical, sibani1993slow, rinn2000multiple,kolvin2012simple,amir2012relaxations, robe2016record, boettcher2018aging, douglass2022distance,kovacs1979isobaric} the microscopic structural mechanisms underlying physical aging remain under debate, and proposed theories are typically built on prior assumptions on the structure of the system's energy landscape, rather than a microscopic mechanism. 

\begin{figure*}
    \centering
    \includegraphics[width=0.85\textwidth]{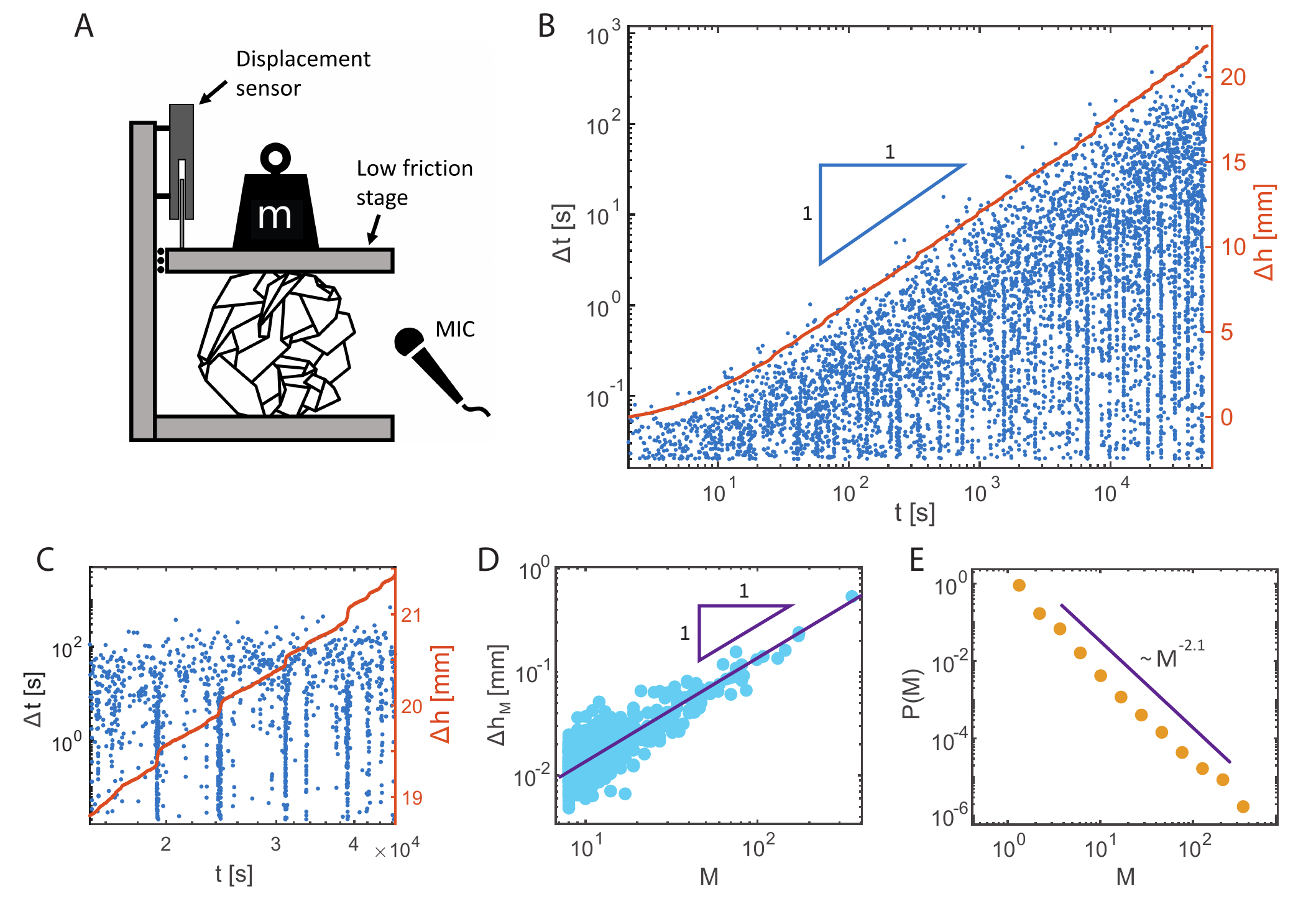}
    \vspace*{-0.2cm}
   \caption{\textbf{Experimental observation of aging and avalanche dynamics} - (\textbf{A}) A schematic illustration of experimental setup. A crumpled sheet is compressed by a load $m$. A displacement sensor measures the compaction $\Delta h$, while a microphone records the acoustic emissions from the sheet; (\textbf{B}) $\Delta h(t)$ (red curve; linear axis on the right) plotted alongside the acoustic activity (blue circles; logarithmic axis on the left). For each acoustic pulse we plot the waiting time since the last pulse $\Delta t$ vs. $t$; (\textbf{C}) Zoom in on panel B, showing correspondence between events grouped to avalanches and sharp features of $h(t)$; (\textbf{D}) Accumulated height in an avalanche $\Delta h_{M}$ vs. avalanche size $M$. The linear scaling indicates each event corresponds to a height drop of $\Delta h_{1}\approx1.4\,\mu$m; (\textbf{E}) Probability distribution of avalanche size $M$, exhibiting a power law decay $P(M)\sim M^{-\alpha}$, with $\alpha=2.1\pm 0.1$.}
    \label{fig:Figure1}
\end{figure*}

Here, through experiments in thin crumpled sheets and simulations of a minimal mechanical model, we reveal the real-space, structural mechanism underlying logarithmic aging in this system. By combining mechanical and acoustic measurements, we show that under constant load, crumpled sheets compact logarithmically via rapid sequences of discrete micro-mechanical instabilities, grouped into avalanches. These avalanches, which we dub \textit{aging avalanches}, have self-similar size and temporal distributions throughout the aging process, yet exhibit timescales and magnitude cut-offs which grow with the system's age. Crucially, we find that the dynamic slow-down of the relaxation process arises from a steady, linear growth of quiescent dwell times between avalanches.

Both logarithmic compaction and the avalanche dynamics are reproduced in a minimal numerical model comprised of a disordered network of bi-stable elastic elements \cite{doi:10.1073/pnas.1300534110, shohat2022memory}, previously used to explain memory formation in crumpled sheets under cyclic drive. The model fully reveals the structural mechanism underlying slow relaxations, intermittency and avalanche dynamics. Our findings can be summarised as follows: under external pressure, the highly frustrated system self-organizes to a state which is poised on the verge of an instability. This localized instability is activated by an otherwise negligible amount of noise. The activation facilitates an avalanche of instabilities, during which new elements are dynamically destabilized and the systems re-organizes such that the avalanche ends when the system is again poised on the verge of an instability, a phenomenology reminiscent of Self Organized Criticality \cite{bak1988self}. Here, however, the magnitude and temporal distributions of the avalanches show aging dynamics, and the quiescent dwell times between avalanches grow with the system’s age due to a small yet steady increase in the lowest effective energy barrier after each avalanche. This leads to a dynamical slow-down, which results in the logarithmic relaxation dynamics.

\textbf{Logarithmic compaction via aging avalanches - } A sheet of Mylar, \(50\,cm\) by \(50\,cm\) across and \(8\,\mu m\) in thickness, is crumpled manually, unfolded and then crumpled again for over 20 times, resulting in a well-developed, complex crease pattern \cite{witten2007stress,blair2005geometry, andrejevic2021model}. After the final crumpling, the sheet is kept in the shape of a loose ball and placed under a freely moving compression stage (see Figure \ref{fig:Figure1}a). Uniaxial load is applied to the system by placing a weight of $m=400$ gr on top of the stage. The displacement of the stage $\Delta h$, representing the vertical compaction of the crumpled ball, is monitored using a high-resolution displacement sensor for over a day. The experimental setup is placed inside an acoustically isolated box, where acoustic emissions from the crumpled sheet are measured at a rate of 48KHz using an amplified microphone (see methods).

\begin{figure*}
    \centering
    \includegraphics[width=1\textwidth]{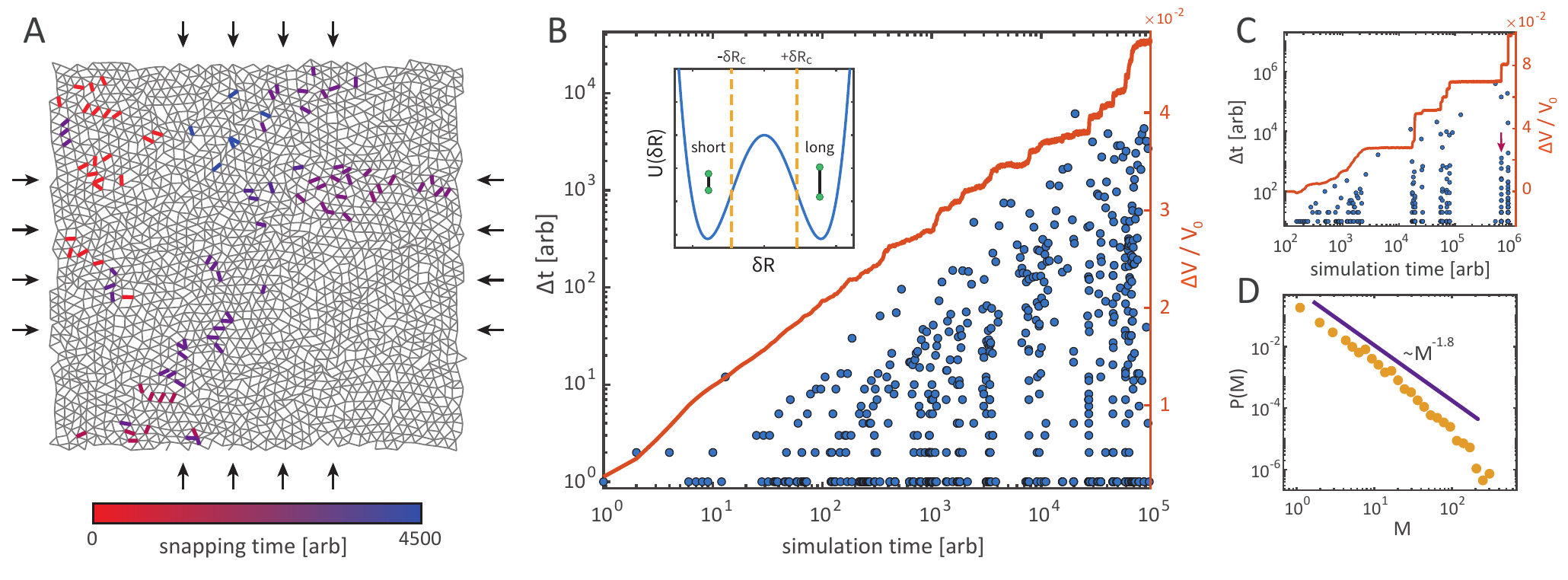}
    \vspace*{-0.6cm}
   \caption{\textbf{Aging in the numerical model} - (\textbf{A}) A simulated disordered network of bi-stable elastic elements with $N=2000$ nodes. The marked bonds participate in a large avalanche marked by an arrow in panel C. Their color represents the intra-avalanche snapping time; (\textbf{B}) Evolution of the normalized compaction $\Delta V/V_{0}$ for a single realization with $N=10^{4}$ (red curve; linear axis on the right), plotted along the waiting times between instability (blue circles; logarithmic axis on the left). Similarly to experiments, instabilities govern the slow compaction of the network. The inset illustrates the bistable potential of a single bond $U(\delta R)$; (\textbf{C}) Normalized compaction $\Delta V/V_{0}$ alongside instabilities for a single realization with $N=2000$, exhibiting highly intermittent activity and a step-wise logarithmic compaction; (\textbf{D}) Probability distribution of avalanche size $P(M)$, collected from ten realization of $N=10^{4}$. The dashed line represents a power-law $P(M)\sim M^{-\alpha}$, with $\alpha=1.8\pm0.2$. }
    \label{fig:Figure2}
\end{figure*}

Under the constant load, the height of the crumpled ball exhibits an ever-slowing logarithmic relaxation \cite{matan2002crumpling} spanning many decades in time, from a second to a day, as shown by the solid red curve in Figure \ref{fig:Figure1}b. During a day of logarithmic relaxation, the crumpled ball emits several thousand acoustic pulses \cite{kramer1996universal,houle1996acoustic,lahini2022Crackling}. These have been shown to originate from the snapping of localized, bi-stable, snap-through instabilities formed across the crumpled sheet, that govern the system's mechanical response \cite{lechenault2015generic, shohat2022memory,shohat2023dissipation}. We record the occurrence time $t_{i}$ of each acoustic pulse during relaxation. 
 
The analysis of the acoustic emission data reveals that the slow relaxation advances via a series of discrete instabilities, many of which are grouped into avalanches. This information is plotted in Fig. \ref{fig:Figure1}b,c, where for each acoustic pulse (blue dot), we plot the waiting time since the previous event $\Delta t_{i}$ on the same time axis as the compaction relaxation data (red curve). Two features are visible in the data: First, the upper cutoff of the typical waiting time between events grows linearly with time, namely $\Delta t/t=C$. This is similar to the results observed in \cite{lahini2022Crackling}, and agrees with the overall logarithmic slow-down of the system's dynamics. Second, and in contrast to previous observations, the system exhibits rich avalanche dynamics. Each avalanche, indicated by a vertical streak in the waiting times plot, consist of many consecutive instabilities, separated by waiting times much shorter than the waiting time cut-off $Ct$. These avalanches are correlated with sharp, step-like features in the compaction curve, as shown in the magnified view in Fig. \ref{fig:Figure1}c. The magnitude of an avalanche $M$, defined by the number of instabilities (see methods and the following paragraph), is proportional to the corresponding height drop $\Delta h_{M}$, as shown in Fig. \ref{fig:Figure1}d. Overall, the cumulative number of instabilities along the experiment traces the overall height relaxation \cite{supplementary}, indicating that the relaxation is governed by the collective dynamics of these instabilities.  

Next, we characterise the avalanche statistics. In contrast to steady-state avalanches observed in slowly driven systems, here the avalanches also show signs of aging - older avalanches span more time-scales, with a cut-off that grows linearly with the age of the system. To account for these growing timescales, we divide the logarithmic time axis, rather than the time itself, to equally spaced bins. The width of the bins follows $w=\langle C\rangle t$, with $\langle C\rangle=\langle \frac{\Delta t}{t}\rangle$ the average normalized waiting time (see methods). We then count the number of instabilities in consecutive non-empty bins \cite{beggs2003neuronal}. This number, denoted $M$, constitutes the avalanche size.  In this division $\sim75\%$ of the bins are empty, indicating that the relaxation is highly intermittent. The avalanche size $M$ is distributed according to a power-law $P(M)\sim M^{-\alpha}$, with $\alpha=2.1\pm 0.1$ (Fig. \ref{fig:Figure1}d). While the range of timescales within avalanches and their magnitude grow with the system's age, we find that the avalanches remain self-similar throughout the aging process, i.e. avalanches from different ages follow the same scaling laws \cite{supplementary}. These results suggest that the system is in a critical state throughout the aging process.

\textbf{Minimal mechanical model - } In previous work, we introduced a minimal numerical model for the quasistatic mechanics of crumpled sheets \cite{shohat2022memory}. The model consists of a disordered two-dimensional network of bi-stable elastic bonds, where each bond has two distinct rest lengths - representing the localized bi-stable degrees of freedom observed experimentally. Under quasi-static drive, this model successfully reproduced several experimentally observed mechanical properties of crumpled sheets, including intermittency, hysteresis, convergence to limit cycles, memory formation, and clear signatures of frustrated interactions between instabilities.

We now show that a dynamic version of this model captures all the experimental results reported here: logarithmic aging and the emergence of long time scales, the crackling avalanche statistics, and the critical behavior. The dynamic model is realized using the molecular dynamics platform LAMMPS \cite{thompson2022lammps} (see methods and SI for a detailed description \cite{supplementary}). The simulation consists of two-dimensional disordered arrangement of $N$ nodes, connected by a network of bi-stable elastic bonds (see Fig \ref{fig:Figure2}a). Each bi-stable bond has a double-well potential of the form $U=\frac{a_{4}}{4}(\delta R)^4-\frac{a_{2}}{2}(\delta R)^2$, where $\delta R=R-R_0$ represents the deviation from the mean of the bond's two rest lengths (see inset in Fig \ref{fig:Figure2}b). The parameters $a_{4}=1$ and $a_{2}=2.5$ are identical for all bonds, while $R_0$ is randomized (see methods). As a result, the different bonds have different rest lengths but an identical energetic barrier separating their two minima, given by $\Delta U_{0}=\frac{a_{2}^2}{4a_{4}}\approx{1.5}$. 

In the first stage of the simulations, we subject the network to constant external pressure, representing the external load, and evolve the system using Langevin dynamics at zero temperature. Under these conditions, the system first contracts under the load and then settles to mechanical equilibrium with a fixed volume $V_{0}$. However, due to incompatibility between the bonds' rest lengths $R_{ij}$, this state is highly frustrated and carries excess stresses, and many bonds deviate from their minimal energy state.

After the system reaches this state, we add noise to the simulations in the form of a very small temperature $T=10^{-3}\ll\Delta U_{0}$ (setting $K_{B}=1$). Namely, $T$ is much lower than the value needed in order to flip any bond between its minima. Nevertheless, the system immediately starts to compact via sequences of instabilities in which many bonds shift to their shorter rest length. The compaction rate decreases in time, resulting in a logarithmic global compaction curve , as shown in Fig. \ref{fig:Figure2}b for $N=10^{4}$. The slow compaction exhibits clear intermittent features, similarly to experiments. Simulating smaller networks with $N=2000$, highlights the intermittent nature of the slow relaxation (Fig. \ref{fig:Figure2}c). Here, activity is strictly grouped into sudden compaction steps, separated by long waiting times in which the system remains stationary, as illustrated in the supplementary movie \cite{supplementary}. These waiting times grow as the system ages, resulting in an ever-slowing, step-wise logarithmic relaxations.

To compare the numerical results to the experimental acoustic measurements, we consider the snapping of individual bonds between their minima to be the equivalent of the snap-through instabilities in the crumpled sheets, which have been measured via their acoustic emission. We then turn to analyze their snapping statistics in a similar manner. The waiting times between individual snaps are plotted alongside the overall compaction in Fig. \ref{fig:Figure2}b. This reveals a scenario similar to the one observed in the experiments: a linear increase in the cut-off of waiting times between events, grouping of events to avalanches, and a correlation between avalanches and discontinuous features in the relaxation curve \cite{supplementary}. Using the model it is possible to resolve the full spatio-temporal dynamics of avalanches, as shown for a typical avalanche in Fig. \ref{fig:Figure2}a.

The avalanche statistical analysis is performed in the same manner as described above for the experiment. Namely, the simulation time is divided into bins that are equally spaces in the logarithm of time, and the number of instabilities $M$ in consecutive non-empty bins is counted. Averaging over many realizations of the dynamics we find that $M$ exhibits a clear power-law distribution $P(M)\sim M^{-\alpha}$, with $\alpha\approx1.8\pm0.2$, as shown in Fig. \ref{fig:Figure2}d. We note that the distribution cutoff grows with system size and that the exponent $\alpha$ depends on the preparation of the network \cite{supplementary}. 

Altogether, these results show that a simple model, a network of coupled instabilities, captures and links all the features of the aging dynamics observed in the experiments. The simplicity of this model offers a unique opportunity to reveal the microscopic structural mechanisms underlying these behaviours, which we describe next.

\begin{figure}
    \includegraphics[width=0.5\textwidth]{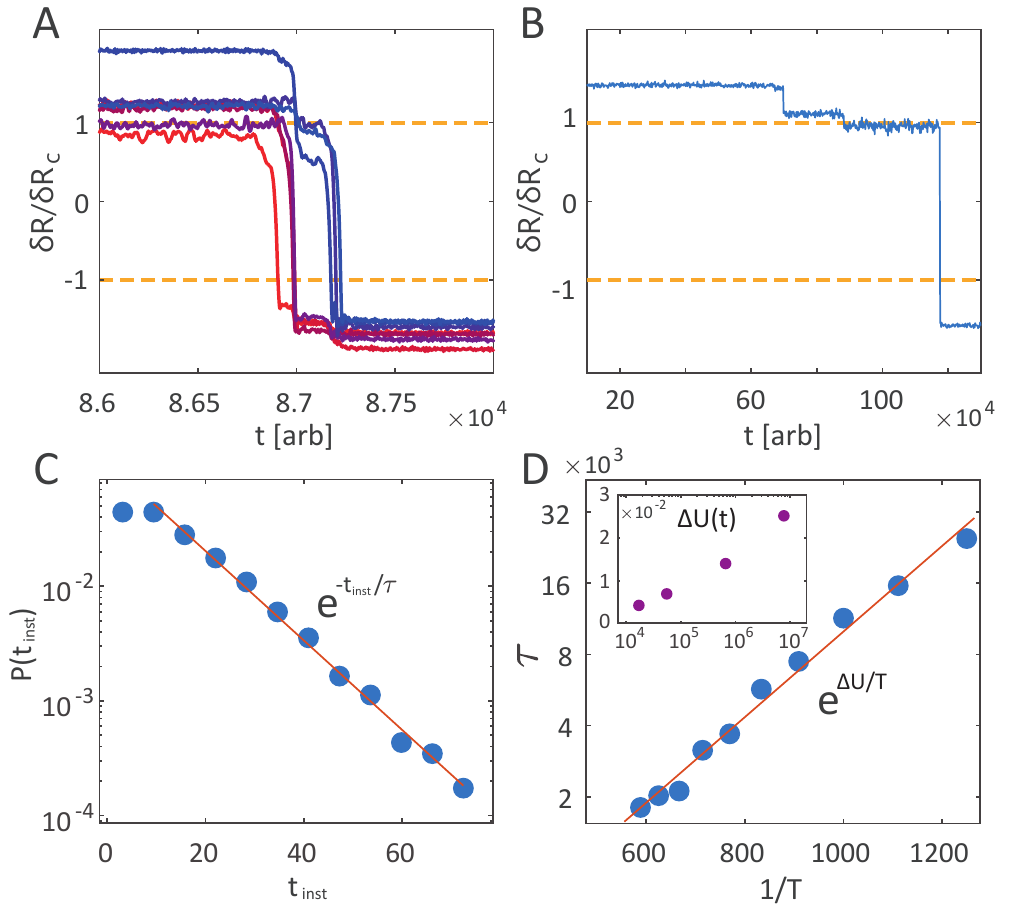}
    \caption{\textbf{Microscopic aging mechanism} - (\textbf{A}) Normalized displacement ${\delta R}/{\delta R_{c}}$ vs $t$ for 6 bonds that comprise a small avalanche. The avalanche is triggered by an unstable bond marked in red;  (\textbf{B}) Evolution of ${\delta R}/{\delta R_{c}}$ for a typical bond. Interactions with other bonds allow it to abruptly approach instability before snapping; (\textbf{C}) $P(t_{inst})$ for a single plateau, exhibiting an exponential tail; (\textbf{D}) Average instability time $\tau$ vs. $1/T$, exhibiting Arrhenius scaling. The inset shows the barriers $\Delta U$ vs. $t_{W}$, the age at the beginning of the plateau before crossing them. The barriers slowly increase with age.}
    \label{fig:Figure3}
\end{figure}

\textbf{Microscopic, real-space mechanism for aging -} The numerical framework described above allows us to study in detail the microscopic properties of the network and its dynamics. We now use this capability to reveal the mechanism underlying the avalanche dynamics, the growth of quiescent dwell times between avalanches and the resulting logarithmic aging. The crux of the mechanism arises from the geometric incompatibility between the rest-lengths of the bonds in the disordered network. Due to this inherent frustration, many of the bonds are stressed, i.e. they deviate from their minima \cite{shohat2022memory}. In particular, the system self-organizes such that some of the bonds lie on the verge of instability. These bonds, we find, play a pivotal role in the aging dynamics. 

As we have shown above, the aging process advanced via discrete compaction steps, separated by long dwell times. Each compaction step corresponds to an avalanche of size $M$, in which many bonds snap in succession. Using high temporal resolution simulations, we find that each avalanche is typically initiated by the snapping of a single bond which lies on the verge of instability. In other words, before snapping this bond lies close to its instability point $\delta R_{c}=\sqrt{\frac{a_{2}}{3a_{4}}}$ which satisfies $U''(\delta R_{c})=0$. One example of an avalanche initiation process is shown in Fig. \ref{fig:Figure3}a in high temporal resolution. The initiating bond, shown in red, lies initially close to its instability point $\delta R_{c}$. After some dwell time in which the system seems to be stationary, this bond suddenly snaps, and interacts with other bonds by redistributing stresses in the network. As a result, some of these bonds destabilize (approach $\delta R_{c}$), and eventually snap, destabilizing other bonds and so forth. We find that avalanches have both inertial and thermal components. Namely, some bonds destabilize and snap even if the temperature is turned off after the first bond snaps, while others are facilitated close to instability yet require thermal noise for snapping. The signature of these facilitation processes is also apparent when tracking the displacement of a single bond over a long time, as shown for example in Fig. \ref{fig:Figure3}b. Before snapping, the bond approaches instability in successive steps, each corresponding to an avalanche occurring somewhere in the system.

Next, we investigate the origin of the long quiescent dwell times between avalanches, signified by the long plateaus in Figure \ref{fig:Figure2}c. Here, no events occur and the system is stationary, aside from equilibrium-like thermal fluctuations \cite{supplementary}. Nevertheless, each plateau ends abruptly with an avalanche. To characterize this process, we focus on a single plateau and consider the \textit{isoconfigurational ensemble} \cite{widmer2004reproducible}. In essence, we run many simulations with the same initial condition (the configuration at the beginning of the plateau) and the same temperature, but with different realizations of the thermal fluctuations. Using this method, it is possible to estimate the dwell time distribution at the plateau. We find that the plateau is always terminated by the same avalanche, while the instability time $t_{inst}$ is exponentially distributed $P(t_{inst})=\frac{1}{\tau} \, exp(- t_{inst}/\tau)$ (Fig. \ref{fig:Figure3}c). For different temperatures, the average instability time $\tau$ exhibits Arrhenius scaling $\tau=\tau_{0}\, exp(\Delta U/T)$ (Fig. \ref{fig:Figure3}d). These results indicate that the instability is triggered via a stochastic first passage process over a \emph{single} energy barrier \cite{hanggi1990reaction}. This barrier can be estimated from the Arrhenius scaling, yielding in our simulations $\Delta U\approx10^{-2}\Delta U_{0}$. 

To investigate the growth of the dwell times during relaxation, we repeat this analysis for several plateaus at different ages of the system. We find that overall, the dynamics of the system reduces to the successive activated crossing of single energy barriers $\Delta U(t)$. However, after each avalanche the network re-organizes and slightly stiffens \cite{supplementary}. As a result, the barrier for the next event increases, as shown in the inset of Fig. \ref{fig:Figure3}d. The scaling $\Delta U\sim log(t)$ is consistent with analytic results for thermal creep in amorphous solids \cite{popovic2022scaling}. The slow increase in barrier height with the system's age accounts for the growing relaxation time scales and the internal slow-down of the dynamics which are central characteristics of aging. 

\textbf{Discussion - } 
Our study reveals the microscopic, real-space mechanism underlying slow relaxation and aging phenomena in a system of coupled bi-stable degrees of freedom. We find that the system self-organizes to a state which is poised on the verge of a local instability. The activation of this instability, corresponding to first passage of a single energy barrier, results in an avalanche, that ends when again the system is poised on the verge of an instability, and the process repeats. This is reminiscent of the phenomenology of Self Organized Criticality (SOC) \cite{bak1988self}, yet with a twist - the range of time scales and magnitudes spanned by an avalanche, and the quiescent waiting times between avalanches grow steadily with the system's age. The slow-down of the dynamics, which is the hallmark of aging, arises from a slow increase in the effective activation barrier after each avalanche.  

The emerging picture, a successive activated crossing of growing energy barriers, is reminiscent of the phenomenology of \textit{traps models} for aging \cite{bouchaud1992weak,rinn2000multiple}. In particular, it is consistent with strain hardening models which were postulated to explain logarithmic aging in crumpled sheets and granular packings \cite{jaeger1989relaxation,matan2002crumpling}. Our findings provide direct evidence for the structural mechanism underlying this process.

On a broader scope, our findings show that the ever-slowing aging dynamics emerges from growing local barriers, rather than from cooperative, non-local dynamics on growing length scales \cite{lubchenko2007theory,ritort2003glassy}. Nevertheless, interaction and facilitation effects result in system spanning, scale-free avalanches. This results in heterogeneous avalanche dynamics, in line with recent simulations of supercooled liquids \cite{chacko2021elastoplasticity}, mesoscopic models of glass formers \cite{kumar2022mapping,ozawa2023elasticity,tahaei2023scaling}, and dense active matter \cite{keta2023intermittent}. This strengthens the connection between the aging mechanism revealed here and elastic models of glasses \cite{dyre2006colloquium}, and raises the question whether it could account for physical aging in different systems with rough energy landscapes \cite{wales1998archetypal,janke2008rugged}, where noise can induce localized transitions between metastable states.

A question which remains open is the source of noise in the experiments. Our results suggest that under load, instabilities in the systems could be triggered by even the slightest perturbations. In the experiment these could be a result of residual environmental noise and mechanical vibrations. Indeed, Matan \textit{et al.} \cite{matan2002crumpling} reported an increase in the logarithmic relaxation rate of crumpled sheets under load when external vibrations were added. Furthermore, while the simulations show that aging can emerge from frustration between elastic elements without requiring material creep, this process may play a role in experiments \cite{jules2020plasticity}.

Finally, as our model coarse-grains the system to mesoscopic instabilities, it resembles an elasto-plastic model (EPM) of amorphous solids \cite{nicolas2018deformation}. Here, however, there are two notable differences: (i) the molecular dynamics implementation allows a full thermodynamic treatment of the model; (ii) interactions in the network are mediated by real-space Langevin dynamics rather than by a generic elastic stress kernel. While a rigorous mapping between EPMs and the bi-stable spring model is needed, the latter may provide additional insight to the complex dynamics of amorphous solids.

\textbf{Acknowledgments - } We are grateful to Daniel Hexner, Shlomi Reuveni, Eran Bouchbinder, Edan Lerner, and Karina Gonz\'alez-Lop\'ez for insightful discussions, and to Yair Shokef, Haim Diamant, and Yohai Bar-Sinai for comments on the manuscript. We also thank Barak Hirshberg for sharing insights on using LAMMPS, and Daniel Hexner for providing the disordered network topologies. This work was supported by the Israel Science Foundation grants 2096/18 and 2117/22. D.S. acknowledges support from the Clore Israel Foundation.

\textbf{Author Contributions - } Y.F. and Y.L. designed and performed experiments, D.S. and Y.L. wrote the LAMMPS model, ran the simulations and analysed the experimental and numerical results. D.S., Y.F and Y.L. wrote the manuscript. 

\textbf{Ethics declarations- } Competing interests: 
The authors declare no competing interests

 \begin{small}  

\section{Methods}

\subsection{Experimental methods}
A sheet of Mylar, 50 by 50 cm across, is cut from an 8 \(\mu \)m thin space blanket. It is then repeatedly crumpled by hand for over 20 times. Next, the crumpled sheet is loosely shaped into a 3D ball, and placed under a compression stage, fixed to a low friction rail. A linear displacement sensor with resolution of \(\sim6\mu\)m is used to measure the vertical displacement $h$ over time, at a rate of 20KHz. A weight of $m=400$ gr is placed on top of the compression stage in order to drive the system with a constant load. Experiments are performed in a sound isolating chamber, with internal walls covered in sound absorbing foam which diminishes internal acoustic reflections and decreases external noise by 10 dB. A high resolution sensitive microphone placed inside the chamber, captures the acoustic emissions at a rate of 48 KHz. The acoustic emission is captured in sync with the displacement sensor data. The crackling noise emitted from the sheet has a typical frequency of 2 KHz and the raw signal is cleaned with a band-pass filter. Instabilities $t_i$ are identified from peaks in the continuous signal. 

\subsection{Aging avalanches characterization}
To measure avalanches of instabilities, we bunch them into periods of correlated activity. Since the upper cutoff of the waiting times grows linearly with system age $\Delta t=Ct$, we consider activity bins which are similarly spaced. We calculate the average $\langle C\rangle=\langle \frac{\Delta t}{t}\rangle$ over all events. We then use $\langle C\rangle$ to construct logarithmic bins (linearly spaced on a logarithmic time axis), such that the bin widths follow $w(t)=\langle C\rangle t$. We define an avalanche of size $M$ as the number of instabilities in consecutive non-empty bins. This division is analogous to the characterization of neuronal avalanches \cite{beggs2003neuronal,priesemann2014spike}, and distinguishes between correlated and uncorrelated instabilities compared to the instantaneous average timescale $\langle C\rangle t$ \cite{supplementary}. In experiments we find $\langle C\rangle=7\times 10^{-4}$, with approximately $75\%$ of the bins are empty of activity.

From the measured avalanches, we assess the power-law distribution exponent $\alpha$ via the maximum likelihood estimator (MLE). For a series of $N$ measurements $x_i$, drawn from a discrete Pareto distribution $P(x)=\frac{x^{-\alpha}}{\zeta(\alpha)}$ (where $\zeta(\alpha)$ is the Riemann zeta function), The MLE $\hat{\alpha}$ satisfies the implicit equation $\frac{1}{N}\sum_{i}log(x_i)+\frac{\zeta'(\hat{\alpha})}{\zeta(\hat{\alpha})}=0$ \cite{bauke2007parameter}.

\subsection{Numerical methods}
We simulate disordered two-dimensional network of bi-stable elastic springs. Each element represents a two-dimensional projection of the localized instabilities observed in experiments \cite{shohat2022memory}. The networks, comprised of $N=10^{3}-10^{4}$ nodes, are over-coordinated, and derived from jammed sphere packings as detailed in \cite{shohat2022memory}. Each bond in the network has a double-well potential of the form
\begin{equation}
\nonumber
    U_{ij}=\frac{a_{4}}{4}(\delta R_{ij})^4-\frac{a_{2}}{2}(\delta R_{ij})^2
\end{equation}
where $\delta R_{ij}$ is the deviation of the bond from its rest length $R_{ij}=R_{ij}^{0}+\delta R_{ij}$. We randomly assign $R_{ij}^{0}\in[9,11]$, while $a_{4}=1$ and $a_{2}=2.5$ are system constants, consistent with previous simulations \cite{shohat2022memory}. From the \textit{same} network topology, \textit{many} realizations can be generated by shuffling the randomized rest lengths $R_{ij}^{0}$. This results in rich yet reproducible dynamics.

We use the molecular dynamics simulations platform LAMMPS \cite{thompson2022lammps}. Brownian dynamics are simulated using a Langevin thermostat with an integration time step $dt=0.1$, using non-periodic boundary conditions. The mass of each node is set to $m=1$. A force $F=0.45$ acts on the boundaries of the network, serving as the external load \cite{supplementary}.

For the data presented in Fig. \ref{fig:Figure3}, we consider 4 plateaus at different ages of the same parent simulation. At each plateau we consider 10 temperatures, with 200 realizations for each temperature. The realizations differ only by the seed of the pseudo-random thermal fluctuations - a method termed the \textit{isoconfigurational ensemble} \cite{widmer2004reproducible}. The errorbars for $\tau$ and $\Delta U$ are smaller than the marker size.

 \end{small}

\textbf{Data availability - } All data needed to evaluate the conclusions are presented in the paper and/or the Supplementary Information. Experimental datasets are available in the supporting information. Data sets generated during the current study are available from the corresponding author upon reasonable request.

\textbf{Code availability - } An example of the LAMMPS simulation script used in this study is available in the supporting information. Other scripts and numerical data are available from the corresponding author upon reasonable request.

\bibliography{bibl.bib}

\begin{thebibliography}{67}
\expandafter\ifx\csname natexlab\endcsname\relax\def\natexlab#1{#1}\fi
\expandafter\ifx\csname bibnamefont\endcsname\relax
  \def\bibnamefont#1{#1}\fi
\expandafter\ifx\csname bibfnamefont\endcsname\relax
  \def\bibfnamefont#1{#1}\fi
\expandafter\ifx\csname citenamefont\endcsname\relax
  \def\citenamefont#1{#1}\fi
\expandafter\ifx\csname url\endcsname\relax
  \def\url#1{\texttt{#1}}\fi
\expandafter\ifx\csname urlprefix\endcsname\relax\def\urlprefix{URL }\fi
\providecommand{\bibinfo}[2]{#2}
\providecommand{\eprint}[2][]{\url{#2}}

\bibitem[{\citenamefont{Matan et~al.}(2002)\citenamefont{Matan, Williams,
  Witten, and Nagel}}]{matan2002crumpling}
\bibinfo{author}{\bibfnamefont{K.}~\bibnamefont{Matan}},
  \bibinfo{author}{\bibfnamefont{R.~B.} \bibnamefont{Williams}},
  \bibinfo{author}{\bibfnamefont{T.~A.} \bibnamefont{Witten}},
  \bibnamefont{and} \bibinfo{author}{\bibfnamefont{S.~R.} \bibnamefont{Nagel}},
  \bibinfo{journal}{Physical Review Letters} \textbf{\bibinfo{volume}{88}},
  \bibinfo{pages}{076101} (\bibinfo{year}{2002}).

\bibitem[{\citenamefont{Lahini et~al.}(2017)\citenamefont{Lahini, Gottesman,
  Amir, and Rubinstein}}]{lahini2017nonmonotonic}
\bibinfo{author}{\bibfnamefont{Y.}~\bibnamefont{Lahini}},
  \bibinfo{author}{\bibfnamefont{O.}~\bibnamefont{Gottesman}},
  \bibinfo{author}{\bibfnamefont{A.}~\bibnamefont{Amir}}, \bibnamefont{and}
  \bibinfo{author}{\bibfnamefont{S.}~\bibnamefont{Rubinstein}},
  \bibinfo{journal}{Physical Review Letters} \textbf{\bibinfo{volume}{118}}
  (\bibinfo{year}{2017}).

\bibitem[{\citenamefont{Kramer and Lobkovsky}(1996)}]{kramer1996universal}
\bibinfo{author}{\bibfnamefont{E.~M.} \bibnamefont{Kramer}} \bibnamefont{and}
  \bibinfo{author}{\bibfnamefont{A.~E.} \bibnamefont{Lobkovsky}},
  \bibinfo{journal}{Physical Review E} \textbf{\bibinfo{volume}{53}},
  \bibinfo{pages}{1465} (\bibinfo{year}{1996}).

\bibitem[{\citenamefont{Shohat et~al.}(2022)\citenamefont{Shohat, Hexner, and
  Lahini}}]{shohat2022memory}
\bibinfo{author}{\bibfnamefont{D.}~\bibnamefont{Shohat}},
  \bibinfo{author}{\bibfnamefont{D.}~\bibnamefont{Hexner}}, \bibnamefont{and}
  \bibinfo{author}{\bibfnamefont{Y.}~\bibnamefont{Lahini}},
  \bibinfo{journal}{Proceedings of the National Academy of Sciences}
  \textbf{\bibinfo{volume}{119}}, \bibinfo{pages}{e2200028119}
  (\bibinfo{year}{2022}).

\bibitem[{\citenamefont{Houle and Sethna}(1996)}]{houle1996acoustic}
\bibinfo{author}{\bibfnamefont{P.~A.} \bibnamefont{Houle}} \bibnamefont{and}
  \bibinfo{author}{\bibfnamefont{J.~P.} \bibnamefont{Sethna}},
  \bibinfo{journal}{Physical Review E} \textbf{\bibinfo{volume}{54}},
  \bibinfo{pages}{278} (\bibinfo{year}{1996}).

\bibitem[{\citenamefont{Keim et~al.}(2019)\citenamefont{Keim, Paulsen,
  Zeravcic, Sastry, and Nagel}}]{keim2019memory}
\bibinfo{author}{\bibfnamefont{N.~C.} \bibnamefont{Keim}},
  \bibinfo{author}{\bibfnamefont{J.~D.} \bibnamefont{Paulsen}},
  \bibinfo{author}{\bibfnamefont{Z.}~\bibnamefont{Zeravcic}},
  \bibinfo{author}{\bibfnamefont{S.}~\bibnamefont{Sastry}}, \bibnamefont{and}
  \bibinfo{author}{\bibfnamefont{S.~R.} \bibnamefont{Nagel}},
  \bibinfo{journal}{Reviews of Modern Physics} \textbf{\bibinfo{volume}{91}},
  \bibinfo{pages}{035002} (\bibinfo{year}{2019}).

\bibitem[{\citenamefont{Oppenheimer and
  Witten}(2015)}]{Oppenheimer2015shapeable}
\bibinfo{author}{\bibfnamefont{N.}~\bibnamefont{Oppenheimer}} \bibnamefont{and}
  \bibinfo{author}{\bibfnamefont{T.~A.} \bibnamefont{Witten}},
  \bibinfo{journal}{Phys. Rev. E} \textbf{\bibinfo{volume}{92}},
  \bibinfo{pages}{052401} (\bibinfo{year}{2015}).

\bibitem[{\citenamefont{Shohat and Lahini}(2023)}]{shohat2023dissipation}
\bibinfo{author}{\bibfnamefont{D.}~\bibnamefont{Shohat}} \bibnamefont{and}
  \bibinfo{author}{\bibfnamefont{Y.}~\bibnamefont{Lahini}},
  \bibinfo{journal}{Physical Review Letters} \textbf{\bibinfo{volume}{130}},
  \bibinfo{pages}{048202} (\bibinfo{year}{2023}).

\bibitem[{\citenamefont{Lahini et~al.}(2023)\citenamefont{Lahini, Rubinstein,
  and Amir}}]{lahini2022Crackling}
\bibinfo{author}{\bibfnamefont{Y.}~\bibnamefont{Lahini}},
  \bibinfo{author}{\bibfnamefont{S.~M.} \bibnamefont{Rubinstein}},
  \bibnamefont{and} \bibinfo{author}{\bibfnamefont{A.}~\bibnamefont{Amir}},
  \bibinfo{journal}{Physical Review Letters} \textbf{\bibinfo{volume}{130}},
  \bibinfo{pages}{258201} (\bibinfo{year}{2023}).

\bibitem[{200(2003)}]{2003SRaN}
\emph{\bibinfo{title}{Slow Relaxations and Nonequilibrium Dynamics in Condensed
  Matter Les Houches Session LXXVII}}, Les Houches - Ecole d'Ete de Physique
  Theorique, 77 (\bibinfo{publisher}{Springer Berlin Heidelberg},
  \bibinfo{year}{2003}).

\bibitem[{\citenamefont{Gurevich and K{\"u}pfer}(1993)}]{gurevich1993time}
\bibinfo{author}{\bibfnamefont{A.}~\bibnamefont{Gurevich}} \bibnamefont{and}
  \bibinfo{author}{\bibfnamefont{H.}~\bibnamefont{K{\"u}pfer}},
  \bibinfo{journal}{Physical Review B} \textbf{\bibinfo{volume}{48}},
  \bibinfo{pages}{6477} (\bibinfo{year}{1993}).

\bibitem[{\citenamefont{Vaknin et~al.}(2000)\citenamefont{Vaknin, Ovadyahu, and
  Pollak}}]{vaknin2000aging}
\bibinfo{author}{\bibfnamefont{A.}~\bibnamefont{Vaknin}},
  \bibinfo{author}{\bibfnamefont{Z.}~\bibnamefont{Ovadyahu}}, \bibnamefont{and}
  \bibinfo{author}{\bibfnamefont{M.}~\bibnamefont{Pollak}},
  \bibinfo{journal}{Physical review letters} \textbf{\bibinfo{volume}{84}},
  \bibinfo{pages}{3402} (\bibinfo{year}{2000}).

\bibitem[{\citenamefont{Grenet et~al.}(2007)\citenamefont{Grenet, Delahaye,
  Sabra, and Gay}}]{grenet2007anomalous}
\bibinfo{author}{\bibfnamefont{T.}~\bibnamefont{Grenet}},
  \bibinfo{author}{\bibfnamefont{J.}~\bibnamefont{Delahaye}},
  \bibinfo{author}{\bibfnamefont{M.}~\bibnamefont{Sabra}}, \bibnamefont{and}
  \bibinfo{author}{\bibfnamefont{F.}~\bibnamefont{Gay}}, \bibinfo{journal}{The
  European Physical Journal B} \textbf{\bibinfo{volume}{56}},
  \bibinfo{pages}{183} (\bibinfo{year}{2007}).

\bibitem[{\citenamefont{Morgan et~al.}(2020)\citenamefont{Morgan, Avinery,
  Rahamim, Beck, and Saleh}}]{morgan2020glassy}
\bibinfo{author}{\bibfnamefont{I.~L.} \bibnamefont{Morgan}},
  \bibinfo{author}{\bibfnamefont{R.}~\bibnamefont{Avinery}},
  \bibinfo{author}{\bibfnamefont{G.}~\bibnamefont{Rahamim}},
  \bibinfo{author}{\bibfnamefont{R.}~\bibnamefont{Beck}}, \bibnamefont{and}
  \bibinfo{author}{\bibfnamefont{O.~A.} \bibnamefont{Saleh}},
  \bibinfo{journal}{Physical Review Letters} \textbf{\bibinfo{volume}{125}},
  \bibinfo{pages}{058001} (\bibinfo{year}{2020}).

\bibitem[{\citenamefont{Lieleg et~al.}(2011)\citenamefont{Lieleg, Kayser,
  Brambilla, Cipelletti, and Bausch}}]{lieleg2011slow}
\bibinfo{author}{\bibfnamefont{O.}~\bibnamefont{Lieleg}},
  \bibinfo{author}{\bibfnamefont{J.}~\bibnamefont{Kayser}},
  \bibinfo{author}{\bibfnamefont{G.}~\bibnamefont{Brambilla}},
  \bibinfo{author}{\bibfnamefont{L.}~\bibnamefont{Cipelletti}},
  \bibnamefont{and} \bibinfo{author}{\bibfnamefont{A.~R.}
  \bibnamefont{Bausch}}, \bibinfo{journal}{Nature materials}
  \textbf{\bibinfo{volume}{10}}, \bibinfo{pages}{236} (\bibinfo{year}{2011}).

\bibitem[{\citenamefont{Kaplan et~al.}(2021)\citenamefont{Kaplan, Reich, Oster,
  Maoz, Levin-Reisman, Ronin, Gefen, Agam, and
  Balaban}}]{kaplan2021observation}
\bibinfo{author}{\bibfnamefont{Y.}~\bibnamefont{Kaplan}},
  \bibinfo{author}{\bibfnamefont{S.}~\bibnamefont{Reich}},
  \bibinfo{author}{\bibfnamefont{E.}~\bibnamefont{Oster}},
  \bibinfo{author}{\bibfnamefont{S.}~\bibnamefont{Maoz}},
  \bibinfo{author}{\bibfnamefont{I.}~\bibnamefont{Levin-Reisman}},
  \bibinfo{author}{\bibfnamefont{I.}~\bibnamefont{Ronin}},
  \bibinfo{author}{\bibfnamefont{O.}~\bibnamefont{Gefen}},
  \bibinfo{author}{\bibfnamefont{O.}~\bibnamefont{Agam}}, \bibnamefont{and}
  \bibinfo{author}{\bibfnamefont{N.~Q.} \bibnamefont{Balaban}},
  \bibinfo{journal}{Nature} \textbf{\bibinfo{volume}{600}},
  \bibinfo{pages}{290} (\bibinfo{year}{2021}).

\bibitem[{\citenamefont{Ben-David et~al.}(2010)\citenamefont{Ben-David,
  Rubinstein, and Fineberg}}]{ben2010slip}
\bibinfo{author}{\bibfnamefont{O.}~\bibnamefont{Ben-David}},
  \bibinfo{author}{\bibfnamefont{S.~M.} \bibnamefont{Rubinstein}},
  \bibnamefont{and} \bibinfo{author}{\bibfnamefont{J.}~\bibnamefont{Fineberg}},
  \bibinfo{journal}{Nature} \textbf{\bibinfo{volume}{463}}, \bibinfo{pages}{76}
  (\bibinfo{year}{2010}).

\bibitem[{\citenamefont{Kaz et~al.}(2012)\citenamefont{Kaz, McGorty, Mani,
  Brenner, and Manoharan}}]{kaz2012physical}
\bibinfo{author}{\bibfnamefont{D.~M.} \bibnamefont{Kaz}},
  \bibinfo{author}{\bibfnamefont{R.}~\bibnamefont{McGorty}},
  \bibinfo{author}{\bibfnamefont{M.}~\bibnamefont{Mani}},
  \bibinfo{author}{\bibfnamefont{M.~P.} \bibnamefont{Brenner}},
  \bibnamefont{and} \bibinfo{author}{\bibfnamefont{V.~N.}
  \bibnamefont{Manoharan}}, \bibinfo{journal}{Nature materials}
  \textbf{\bibinfo{volume}{11}}, \bibinfo{pages}{138} (\bibinfo{year}{2012}).

\bibitem[{\citenamefont{Song et~al.}(2022)\citenamefont{Song, Zhang,
  de~Quesada, Rizvi, Tracy, Ilavsky, Narayanan, Del~Gado, Leheny,
  Holten-Andersen et~al.}}]{song2022gels}
\bibinfo{author}{\bibfnamefont{J.}~\bibnamefont{Song}},
  \bibinfo{author}{\bibfnamefont{Q.}~\bibnamefont{Zhang}},
  \bibinfo{author}{\bibfnamefont{F.}~\bibnamefont{de~Quesada}},
  \bibinfo{author}{\bibfnamefont{M.~H.} \bibnamefont{Rizvi}},
  \bibinfo{author}{\bibfnamefont{J.~B.} \bibnamefont{Tracy}},
  \bibinfo{author}{\bibfnamefont{J.}~\bibnamefont{Ilavsky}},
  \bibinfo{author}{\bibfnamefont{S.}~\bibnamefont{Narayanan}},
  \bibinfo{author}{\bibfnamefont{E.}~\bibnamefont{Del~Gado}},
  \bibinfo{author}{\bibfnamefont{R.~L.} \bibnamefont{Leheny}},
  \bibinfo{author}{\bibfnamefont{N.}~\bibnamefont{Holten-Andersen}},
  \bibnamefont{et~al.}, \bibinfo{journal}{Proceedings of the National Academy
  of Sciences} \textbf{\bibinfo{volume}{119}}, \bibinfo{pages}{e2201566119}
  (\bibinfo{year}{2022}).

\bibitem[{\citenamefont{Knight et~al.}(1995)\citenamefont{Knight, Fandrich,
  Lau, Jaeger, and Nagel}}]{knight1995density}
\bibinfo{author}{\bibfnamefont{J.~B.} \bibnamefont{Knight}},
  \bibinfo{author}{\bibfnamefont{C.~G.} \bibnamefont{Fandrich}},
  \bibinfo{author}{\bibfnamefont{C.~N.} \bibnamefont{Lau}},
  \bibinfo{author}{\bibfnamefont{H.~M.} \bibnamefont{Jaeger}},
  \bibnamefont{and} \bibinfo{author}{\bibfnamefont{S.~R.} \bibnamefont{Nagel}},
  \bibinfo{journal}{Physical review E} \textbf{\bibinfo{volume}{51}},
  \bibinfo{pages}{3957} (\bibinfo{year}{1995}).

\bibitem[{\citenamefont{Angell}(1995)}]{angell1995formation}
\bibinfo{author}{\bibfnamefont{C.~A.} \bibnamefont{Angell}},
  \bibinfo{journal}{Science} \textbf{\bibinfo{volume}{267}},
  \bibinfo{pages}{1924} (\bibinfo{year}{1995}).

\bibitem[{\citenamefont{Debenedetti and
  Stillinger}(2001)}]{debenedetti2001supercooled}
\bibinfo{author}{\bibfnamefont{P.~G.} \bibnamefont{Debenedetti}}
  \bibnamefont{and} \bibinfo{author}{\bibfnamefont{F.~H.}
  \bibnamefont{Stillinger}}, \bibinfo{journal}{Nature}
  \textbf{\bibinfo{volume}{410}}, \bibinfo{pages}{259} (\bibinfo{year}{2001}).

\bibitem[{\citenamefont{Arceri et~al.}(2020)\citenamefont{Arceri, Landes,
  Berthier, and Biroli}}]{Arceri2020}
\bibinfo{author}{\bibfnamefont{F.}~\bibnamefont{Arceri}},
  \bibinfo{author}{\bibfnamefont{F.~P.} \bibnamefont{Landes}},
  \bibinfo{author}{\bibfnamefont{L.}~\bibnamefont{Berthier}}, \bibnamefont{and}
  \bibinfo{author}{\bibfnamefont{G.}~\bibnamefont{Biroli}},
  \emph{\bibinfo{title}{A Statistical Mechanics Perspective on Glasses and
  Aging}} (\bibinfo{publisher}{Springer Berlin Heidelberg},
  \bibinfo{address}{Berlin, Heidelberg}, \bibinfo{year}{2020}), pp.
  \bibinfo{pages}{1--68}, ISBN \bibinfo{isbn}{978-3-642-27737-5}.

\bibitem[{\citenamefont{Lundgren et~al.}(1983)\citenamefont{Lundgren,
  Svedlindh, Nordblad, and Beckman}}]{lundgren1983dynamics}
\bibinfo{author}{\bibfnamefont{L.}~\bibnamefont{Lundgren}},
  \bibinfo{author}{\bibfnamefont{P.}~\bibnamefont{Svedlindh}},
  \bibinfo{author}{\bibfnamefont{P.}~\bibnamefont{Nordblad}}, \bibnamefont{and}
  \bibinfo{author}{\bibfnamefont{O.}~\bibnamefont{Beckman}},
  \bibinfo{journal}{Physical review letters} \textbf{\bibinfo{volume}{51}},
  \bibinfo{pages}{911} (\bibinfo{year}{1983}).

\bibitem[{\citenamefont{Struik}(1977)}]{struik1977physical}
\bibinfo{author}{\bibfnamefont{L.~C.~E.} \bibnamefont{Struik}},
  \bibinfo{journal}{Polymer Engineering \& Science}
  \textbf{\bibinfo{volume}{17}}, \bibinfo{pages}{165} (\bibinfo{year}{1977}).

\bibitem[{\citenamefont{Qiao and Pelletier}(2014)}]{qiao2014dynamic}
\bibinfo{author}{\bibfnamefont{J.}~\bibnamefont{Qiao}} \bibnamefont{and}
  \bibinfo{author}{\bibfnamefont{J.-M.} \bibnamefont{Pelletier}},
  \bibinfo{journal}{Journal of Materials Science \& Technology}
  \textbf{\bibinfo{volume}{30}}, \bibinfo{pages}{523} (\bibinfo{year}{2014}).

\bibitem[{\citenamefont{Weeks et~al.}(2000)\citenamefont{Weeks, Crocker,
  Levitt, Schofield, and Weitz}}]{weeks2000three}
\bibinfo{author}{\bibfnamefont{E.~R.} \bibnamefont{Weeks}},
  \bibinfo{author}{\bibfnamefont{J.~C.} \bibnamefont{Crocker}},
  \bibinfo{author}{\bibfnamefont{A.~C.} \bibnamefont{Levitt}},
  \bibinfo{author}{\bibfnamefont{A.}~\bibnamefont{Schofield}},
  \bibnamefont{and} \bibinfo{author}{\bibfnamefont{D.~A.} \bibnamefont{Weitz}},
  \bibinfo{journal}{Science} \textbf{\bibinfo{volume}{287}},
  \bibinfo{pages}{627} (\bibinfo{year}{2000}).

\bibitem[{\citenamefont{Samarakoon et~al.}(2016)\citenamefont{Samarakoon, Sato,
  Chen, Chern, Yang, Klich, Sinclair, Zhou, and Lee}}]{samarakoon2016aging}
\bibinfo{author}{\bibfnamefont{A.}~\bibnamefont{Samarakoon}},
  \bibinfo{author}{\bibfnamefont{T.~J.} \bibnamefont{Sato}},
  \bibinfo{author}{\bibfnamefont{T.}~\bibnamefont{Chen}},
  \bibinfo{author}{\bibfnamefont{G.-W.} \bibnamefont{Chern}},
  \bibinfo{author}{\bibfnamefont{J.}~\bibnamefont{Yang}},
  \bibinfo{author}{\bibfnamefont{I.}~\bibnamefont{Klich}},
  \bibinfo{author}{\bibfnamefont{R.}~\bibnamefont{Sinclair}},
  \bibinfo{author}{\bibfnamefont{H.}~\bibnamefont{Zhou}}, \bibnamefont{and}
  \bibinfo{author}{\bibfnamefont{S.-H.} \bibnamefont{Lee}},
  \bibinfo{journal}{Proceedings of the National Academy of Sciences}
  \textbf{\bibinfo{volume}{113}}, \bibinfo{pages}{11806}
  (\bibinfo{year}{2016}).

\bibitem[{\citenamefont{Narayanaswamy}(1971)}]{narayanaswamy1971model}
\bibinfo{author}{\bibfnamefont{O.}~\bibnamefont{Narayanaswamy}},
  \bibinfo{journal}{Journal of the American Ceramic Society}
  \textbf{\bibinfo{volume}{54}}, \bibinfo{pages}{491} (\bibinfo{year}{1971}).

\bibitem[{\citenamefont{Bouchaud}(1992)}]{bouchaud1992weak}
\bibinfo{author}{\bibfnamefont{J.-P.} \bibnamefont{Bouchaud}},
  \bibinfo{journal}{Journal de Physique I} \textbf{\bibinfo{volume}{2}},
  \bibinfo{pages}{1705} (\bibinfo{year}{1992}).

\bibitem[{\citenamefont{Cugliandolo and
  Kurchan}(1993)}]{cugliandolo1993analytical}
\bibinfo{author}{\bibfnamefont{L.~F.} \bibnamefont{Cugliandolo}}
  \bibnamefont{and} \bibinfo{author}{\bibfnamefont{J.}~\bibnamefont{Kurchan}},
  \bibinfo{journal}{Physical Review Letters} \textbf{\bibinfo{volume}{71}},
  \bibinfo{pages}{173} (\bibinfo{year}{1993}).

\bibitem[{\citenamefont{Sibani and Hoffmann}(1989)}]{sibani1989hierarchical}
\bibinfo{author}{\bibfnamefont{P.}~\bibnamefont{Sibani}} \bibnamefont{and}
  \bibinfo{author}{\bibfnamefont{K.~H.} \bibnamefont{Hoffmann}},
  \bibinfo{journal}{Physical review letters} \textbf{\bibinfo{volume}{63}},
  \bibinfo{pages}{2853} (\bibinfo{year}{1989}).

\bibitem[{\citenamefont{Sibani and Littlewood}(1993)}]{sibani1993slow}
\bibinfo{author}{\bibfnamefont{P.}~\bibnamefont{Sibani}} \bibnamefont{and}
  \bibinfo{author}{\bibfnamefont{P.~B.} \bibnamefont{Littlewood}},
  \bibinfo{journal}{Physical review letters} \textbf{\bibinfo{volume}{71}},
  \bibinfo{pages}{1482} (\bibinfo{year}{1993}).

\bibitem[{\citenamefont{Rinn et~al.}(2000)\citenamefont{Rinn, Maass, and
  Bouchaud}}]{rinn2000multiple}
\bibinfo{author}{\bibfnamefont{B.}~\bibnamefont{Rinn}},
  \bibinfo{author}{\bibfnamefont{P.}~\bibnamefont{Maass}}, \bibnamefont{and}
  \bibinfo{author}{\bibfnamefont{J.-P.} \bibnamefont{Bouchaud}},
  \bibinfo{journal}{Physical review letters} \textbf{\bibinfo{volume}{84}},
  \bibinfo{pages}{5403} (\bibinfo{year}{2000}).

\bibitem[{\citenamefont{Kolvin and Bouchbinder}(2012)}]{kolvin2012simple}
\bibinfo{author}{\bibfnamefont{I.}~\bibnamefont{Kolvin}} \bibnamefont{and}
  \bibinfo{author}{\bibfnamefont{E.}~\bibnamefont{Bouchbinder}},
  \bibinfo{journal}{Physical Review E} \textbf{\bibinfo{volume}{86}},
  \bibinfo{pages}{010501} (\bibinfo{year}{2012}).

\bibitem[{\citenamefont{Amir et~al.}(2012)\citenamefont{Amir, Oreg, and
  Imry}}]{amir2012relaxations}
\bibinfo{author}{\bibfnamefont{A.}~\bibnamefont{Amir}},
  \bibinfo{author}{\bibfnamefont{Y.}~\bibnamefont{Oreg}}, \bibnamefont{and}
  \bibinfo{author}{\bibfnamefont{Y.}~\bibnamefont{Imry}},
  \bibinfo{journal}{Proceedings of the National Academy of Sciences}
  \textbf{\bibinfo{volume}{109}}, \bibinfo{pages}{1850} (\bibinfo{year}{2012}).

\bibitem[{\citenamefont{Robe et~al.}(2016)\citenamefont{Robe, Boettcher,
  Sibani, and Yunker}}]{robe2016record}
\bibinfo{author}{\bibfnamefont{D.~M.} \bibnamefont{Robe}},
  \bibinfo{author}{\bibfnamefont{S.}~\bibnamefont{Boettcher}},
  \bibinfo{author}{\bibfnamefont{P.}~\bibnamefont{Sibani}}, \bibnamefont{and}
  \bibinfo{author}{\bibfnamefont{P.}~\bibnamefont{Yunker}},
  \bibinfo{journal}{EPL (Europhysics Letters)} \textbf{\bibinfo{volume}{116}},
  \bibinfo{pages}{38003} (\bibinfo{year}{2016}).

\bibitem[{\citenamefont{Boettcher et~al.}(2018)\citenamefont{Boettcher, Robe,
  and Sibani}}]{boettcher2018aging}
\bibinfo{author}{\bibfnamefont{S.}~\bibnamefont{Boettcher}},
  \bibinfo{author}{\bibfnamefont{D.~M.} \bibnamefont{Robe}}, \bibnamefont{and}
  \bibinfo{author}{\bibfnamefont{P.}~\bibnamefont{Sibani}},
  \bibinfo{journal}{Physical Review E} \textbf{\bibinfo{volume}{98}},
  \bibinfo{pages}{020602} (\bibinfo{year}{2018}).

\bibitem[{\citenamefont{Douglass and Dyre}(2022)}]{douglass2022distance}
\bibinfo{author}{\bibfnamefont{I.~M.} \bibnamefont{Douglass}} \bibnamefont{and}
  \bibinfo{author}{\bibfnamefont{J.~C.} \bibnamefont{Dyre}},
  \bibinfo{journal}{Physical Review E} \textbf{\bibinfo{volume}{106}},
  \bibinfo{pages}{054615} (\bibinfo{year}{2022}).

\bibitem[{\citenamefont{Kovacs et~al.}(1979)\citenamefont{Kovacs, Aklonis,
  Hutchinson, and Ramos}}]{kovacs1979isobaric}
\bibinfo{author}{\bibfnamefont{A.~J.} \bibnamefont{Kovacs}},
  \bibinfo{author}{\bibfnamefont{J.~J.} \bibnamefont{Aklonis}},
  \bibinfo{author}{\bibfnamefont{J.~M.} \bibnamefont{Hutchinson}},
  \bibnamefont{and} \bibinfo{author}{\bibfnamefont{A.~R.} \bibnamefont{Ramos}},
  \bibinfo{journal}{Journal of Polymer Science: Polymer Physics Edition}
  \textbf{\bibinfo{volume}{17}}, \bibinfo{pages}{1097} (\bibinfo{year}{1979}).

\bibitem[{\citenamefont{Yan et~al.}(2013)\citenamefont{Yan, Düring, and
  Wyart}}]{doi:10.1073/pnas.1300534110}
\bibinfo{author}{\bibfnamefont{L.}~\bibnamefont{Yan}},
  \bibinfo{author}{\bibfnamefont{G.}~\bibnamefont{Düring}}, \bibnamefont{and}
  \bibinfo{author}{\bibfnamefont{M.}~\bibnamefont{Wyart}},
  \bibinfo{journal}{Proceedings of the National Academy of Sciences}
  \textbf{\bibinfo{volume}{110}}, \bibinfo{pages}{6307} (\bibinfo{year}{2013}).

\bibitem[{\citenamefont{Bak et~al.}(1988)\citenamefont{Bak, Tang, and
  Wiesenfeld}}]{bak1988self}
\bibinfo{author}{\bibfnamefont{P.}~\bibnamefont{Bak}},
  \bibinfo{author}{\bibfnamefont{C.}~\bibnamefont{Tang}}, \bibnamefont{and}
  \bibinfo{author}{\bibfnamefont{K.}~\bibnamefont{Wiesenfeld}},
  \bibinfo{journal}{Physical review A} \textbf{\bibinfo{volume}{38}},
  \bibinfo{pages}{364} (\bibinfo{year}{1988}).

\bibitem[{\citenamefont{Witten}(2007)}]{witten2007stress}
\bibinfo{author}{\bibfnamefont{T.~A.} \bibnamefont{Witten}},
  \bibinfo{journal}{Reviews of Modern Physics} \textbf{\bibinfo{volume}{79}},
  \bibinfo{pages}{643} (\bibinfo{year}{2007}).

\bibitem[{\citenamefont{Blair and Kudrolli}(2005)}]{blair2005geometry}
\bibinfo{author}{\bibfnamefont{D.~L.} \bibnamefont{Blair}} \bibnamefont{and}
  \bibinfo{author}{\bibfnamefont{A.}~\bibnamefont{Kudrolli}},
  \bibinfo{journal}{Physical review letters} \textbf{\bibinfo{volume}{94}},
  \bibinfo{pages}{166107} (\bibinfo{year}{2005}).

\bibitem[{\citenamefont{Andrejevic et~al.}(2021)\citenamefont{Andrejevic, Lee,
  Rubinstein, and Rycroft}}]{andrejevic2021model}
\bibinfo{author}{\bibfnamefont{J.}~\bibnamefont{Andrejevic}},
  \bibinfo{author}{\bibfnamefont{L.~M.} \bibnamefont{Lee}},
  \bibinfo{author}{\bibfnamefont{S.~M.} \bibnamefont{Rubinstein}},
  \bibnamefont{and} \bibinfo{author}{\bibfnamefont{C.~H.}
  \bibnamefont{Rycroft}}, \bibinfo{journal}{Nature communications}
  \textbf{\bibinfo{volume}{12}}, \bibinfo{pages}{1} (\bibinfo{year}{2021}).

\bibitem[{\citenamefont{Lechenault and
  Adda-Bedia}(2015)}]{lechenault2015generic}
\bibinfo{author}{\bibfnamefont{F.}~\bibnamefont{Lechenault}} \bibnamefont{and}
  \bibinfo{author}{\bibfnamefont{M.}~\bibnamefont{Adda-Bedia}},
  \bibinfo{journal}{Physical review letters} \textbf{\bibinfo{volume}{115}},
  \bibinfo{pages}{235501} (\bibinfo{year}{2015}).

\bibitem[{sup()}]{supplementary}
\bibinfo{note}{See supplementary information for further details}.

\bibitem[{\citenamefont{Beggs and Plenz}(2003)}]{beggs2003neuronal}
\bibinfo{author}{\bibfnamefont{J.~M.} \bibnamefont{Beggs}} \bibnamefont{and}
  \bibinfo{author}{\bibfnamefont{D.}~\bibnamefont{Plenz}},
  \bibinfo{journal}{Journal of neuroscience} \textbf{\bibinfo{volume}{23}},
  \bibinfo{pages}{11167} (\bibinfo{year}{2003}).

\bibitem[{\citenamefont{Thompson et~al.}(2022)\citenamefont{Thompson, Aktulga,
  Berger, Bolintineanu, Brown, Crozier, in't Veld, Kohlmeyer, Moore, Nguyen
  et~al.}}]{thompson2022lammps}
\bibinfo{author}{\bibfnamefont{A.~P.} \bibnamefont{Thompson}},
  \bibinfo{author}{\bibfnamefont{H.~M.} \bibnamefont{Aktulga}},
  \bibinfo{author}{\bibfnamefont{R.}~\bibnamefont{Berger}},
  \bibinfo{author}{\bibfnamefont{D.~S.} \bibnamefont{Bolintineanu}},
  \bibinfo{author}{\bibfnamefont{W.~M.} \bibnamefont{Brown}},
  \bibinfo{author}{\bibfnamefont{P.~S.} \bibnamefont{Crozier}},
  \bibinfo{author}{\bibfnamefont{P.~J.} \bibnamefont{in't Veld}},
  \bibinfo{author}{\bibfnamefont{A.}~\bibnamefont{Kohlmeyer}},
  \bibinfo{author}{\bibfnamefont{S.~G.} \bibnamefont{Moore}},
  \bibinfo{author}{\bibfnamefont{T.~D.} \bibnamefont{Nguyen}},
  \bibnamefont{et~al.}, \bibinfo{journal}{Computer Physics Communications}
  \textbf{\bibinfo{volume}{271}}, \bibinfo{pages}{108171}
  (\bibinfo{year}{2022}).

\bibitem[{\citenamefont{Widmer-Cooper et~al.}(2004)\citenamefont{Widmer-Cooper,
  Harrowell, and Fynewever}}]{widmer2004reproducible}
\bibinfo{author}{\bibfnamefont{A.}~\bibnamefont{Widmer-Cooper}},
  \bibinfo{author}{\bibfnamefont{P.}~\bibnamefont{Harrowell}},
  \bibnamefont{and}
  \bibinfo{author}{\bibfnamefont{H.}~\bibnamefont{Fynewever}},
  \bibinfo{journal}{Physical review letters} \textbf{\bibinfo{volume}{93}},
  \bibinfo{pages}{135701} (\bibinfo{year}{2004}).

\bibitem[{\citenamefont{H{\"a}nggi et~al.}(1990)\citenamefont{H{\"a}nggi,
  Talkner, and Borkovec}}]{hanggi1990reaction}
\bibinfo{author}{\bibfnamefont{P.}~\bibnamefont{H{\"a}nggi}},
  \bibinfo{author}{\bibfnamefont{P.}~\bibnamefont{Talkner}}, \bibnamefont{and}
  \bibinfo{author}{\bibfnamefont{M.}~\bibnamefont{Borkovec}},
  \bibinfo{journal}{Reviews of modern physics} \textbf{\bibinfo{volume}{62}},
  \bibinfo{pages}{251} (\bibinfo{year}{1990}).

\bibitem[{\citenamefont{Popovic et~al.}(2022)\citenamefont{Popovic, de~Geus,
  Ji, Rosso, and Wyart}}]{popovic2022scaling}
\bibinfo{author}{\bibfnamefont{M.}~\bibnamefont{Popovic}},
  \bibinfo{author}{\bibfnamefont{T.~W.} \bibnamefont{de~Geus}},
  \bibinfo{author}{\bibfnamefont{W.}~\bibnamefont{Ji}},
  \bibinfo{author}{\bibfnamefont{A.}~\bibnamefont{Rosso}}, \bibnamefont{and}
  \bibinfo{author}{\bibfnamefont{M.}~\bibnamefont{Wyart}},
  \bibinfo{journal}{Physical Review Letters} \textbf{\bibinfo{volume}{129}}
  (\bibinfo{year}{2022}).

\bibitem[{\citenamefont{Jaeger et~al.}(1989)\citenamefont{Jaeger, Liu, and
  Nagel}}]{jaeger1989relaxation}
\bibinfo{author}{\bibfnamefont{H.}~\bibnamefont{Jaeger}},
  \bibinfo{author}{\bibfnamefont{C.-h.} \bibnamefont{Liu}}, \bibnamefont{and}
  \bibinfo{author}{\bibfnamefont{S.~R.} \bibnamefont{Nagel}},
  \bibinfo{journal}{Physical Review Letters} \textbf{\bibinfo{volume}{62}},
  \bibinfo{pages}{40} (\bibinfo{year}{1989}).

\bibitem[{\citenamefont{Lubchenko and Wolynes}(2007)}]{lubchenko2007theory}
\bibinfo{author}{\bibfnamefont{V.}~\bibnamefont{Lubchenko}} \bibnamefont{and}
  \bibinfo{author}{\bibfnamefont{P.~G.} \bibnamefont{Wolynes}},
  \bibinfo{journal}{Annu. Rev. Phys. Chem.} \textbf{\bibinfo{volume}{58}},
  \bibinfo{pages}{235} (\bibinfo{year}{2007}).

\bibitem[{\citenamefont{Ritort and Sollich}(2003)}]{ritort2003glassy}
\bibinfo{author}{\bibfnamefont{F.}~\bibnamefont{Ritort}} \bibnamefont{and}
  \bibinfo{author}{\bibfnamefont{P.}~\bibnamefont{Sollich}},
  \bibinfo{journal}{Advances in physics} \textbf{\bibinfo{volume}{52}},
  \bibinfo{pages}{219} (\bibinfo{year}{2003}).

\bibitem[{\citenamefont{Chacko et~al.}(2021)\citenamefont{Chacko, Landes,
  Biroli, Dauchot, Liu, and Reichman}}]{chacko2021elastoplasticity}
\bibinfo{author}{\bibfnamefont{R.~N.} \bibnamefont{Chacko}},
  \bibinfo{author}{\bibfnamefont{F.~P.} \bibnamefont{Landes}},
  \bibinfo{author}{\bibfnamefont{G.}~\bibnamefont{Biroli}},
  \bibinfo{author}{\bibfnamefont{O.}~\bibnamefont{Dauchot}},
  \bibinfo{author}{\bibfnamefont{A.~J.} \bibnamefont{Liu}}, \bibnamefont{and}
  \bibinfo{author}{\bibfnamefont{D.~R.} \bibnamefont{Reichman}},
  \bibinfo{journal}{Physical Review Letters} \textbf{\bibinfo{volume}{127}},
  \bibinfo{pages}{048002} (\bibinfo{year}{2021}).

\bibitem[{\citenamefont{Kumar et~al.}(2022)\citenamefont{Kumar, Patinet,
  Maloney, Regev, Vandembroucq, and Mungan}}]{kumar2022mapping}
\bibinfo{author}{\bibfnamefont{D.}~\bibnamefont{Kumar}},
  \bibinfo{author}{\bibfnamefont{S.}~\bibnamefont{Patinet}},
  \bibinfo{author}{\bibfnamefont{C.~E.} \bibnamefont{Maloney}},
  \bibinfo{author}{\bibfnamefont{I.}~\bibnamefont{Regev}},
  \bibinfo{author}{\bibfnamefont{D.}~\bibnamefont{Vandembroucq}},
  \bibnamefont{and} \bibinfo{author}{\bibfnamefont{M.}~\bibnamefont{Mungan}},
  \bibinfo{journal}{The Journal of Chemical Physics}
  \textbf{\bibinfo{volume}{157}}, \bibinfo{pages}{174504}
  (\bibinfo{year}{2022}).

\bibitem[{\citenamefont{Ozawa and Biroli}(2023)}]{ozawa2023elasticity}
\bibinfo{author}{\bibfnamefont{M.}~\bibnamefont{Ozawa}} \bibnamefont{and}
  \bibinfo{author}{\bibfnamefont{G.}~\bibnamefont{Biroli}},
  \bibinfo{journal}{Physical Review Letters} \textbf{\bibinfo{volume}{130}},
  \bibinfo{pages}{138201} (\bibinfo{year}{2023}).

\bibitem[{\citenamefont{Tahaei et~al.}(2023)\citenamefont{Tahaei, Biroli,
  Ozawa, Popovi{\'c}, and Wyart}}]{tahaei2023scaling}
\bibinfo{author}{\bibfnamefont{A.}~\bibnamefont{Tahaei}},
  \bibinfo{author}{\bibfnamefont{G.}~\bibnamefont{Biroli}},
  \bibinfo{author}{\bibfnamefont{M.}~\bibnamefont{Ozawa}},
  \bibinfo{author}{\bibfnamefont{M.}~\bibnamefont{Popovi{\'c}}},
  \bibnamefont{and} \bibinfo{author}{\bibfnamefont{M.}~\bibnamefont{Wyart}},
  \bibinfo{journal}{arXiv preprint arXiv:2305.00219}  (\bibinfo{year}{2023}).

\bibitem[{\citenamefont{Keta et~al.}(2023)\citenamefont{Keta, Mandal, Sollich,
  Jack, and Berthier}}]{keta2023intermittent}
\bibinfo{author}{\bibfnamefont{Y.-E.} \bibnamefont{Keta}},
  \bibinfo{author}{\bibfnamefont{R.}~\bibnamefont{Mandal}},
  \bibinfo{author}{\bibfnamefont{P.}~\bibnamefont{Sollich}},
  \bibinfo{author}{\bibfnamefont{R.~L.} \bibnamefont{Jack}}, \bibnamefont{and}
  \bibinfo{author}{\bibfnamefont{L.}~\bibnamefont{Berthier}},
  \bibinfo{journal}{Soft Matter}  (\bibinfo{year}{2023}).

\bibitem[{\citenamefont{Dyre}(2006)}]{dyre2006colloquium}
\bibinfo{author}{\bibfnamefont{J.~C.} \bibnamefont{Dyre}},
  \bibinfo{journal}{Reviews of modern physics} \textbf{\bibinfo{volume}{78}},
  \bibinfo{pages}{953} (\bibinfo{year}{2006}).

\bibitem[{\citenamefont{Wales et~al.}(1998)\citenamefont{Wales, Miller, and
  Walsh}}]{wales1998archetypal}
\bibinfo{author}{\bibfnamefont{D.~J.} \bibnamefont{Wales}},
  \bibinfo{author}{\bibfnamefont{M.~A.} \bibnamefont{Miller}},
  \bibnamefont{and} \bibinfo{author}{\bibfnamefont{T.~R.} \bibnamefont{Walsh}},
  \bibinfo{journal}{Nature} \textbf{\bibinfo{volume}{394}},
  \bibinfo{pages}{758} (\bibinfo{year}{1998}).

\bibitem[{\citenamefont{Janke and Janke}(2008)}]{janke2008rugged}
\bibinfo{author}{\bibfnamefont{W.}~\bibnamefont{Janke}} \bibnamefont{and}
  \bibinfo{author}{\bibfnamefont{W.}~\bibnamefont{Janke}},
  \emph{\bibinfo{title}{Rugged Free-Energy Landscapes--An Introduction}}
  (\bibinfo{publisher}{Springer}, \bibinfo{year}{2008}).

\bibitem[{\citenamefont{Jules et~al.}(2020)\citenamefont{Jules, Lechenault, and
  Adda-Bedia}}]{jules2020plasticity}
\bibinfo{author}{\bibfnamefont{T.}~\bibnamefont{Jules}},
  \bibinfo{author}{\bibfnamefont{F.}~\bibnamefont{Lechenault}},
  \bibnamefont{and}
  \bibinfo{author}{\bibfnamefont{M.}~\bibnamefont{Adda-Bedia}},
  \bibinfo{journal}{Physical Review E} \textbf{\bibinfo{volume}{102}},
  \bibinfo{pages}{033005} (\bibinfo{year}{2020}).

\bibitem[{\citenamefont{Nicolas et~al.}(2018)\citenamefont{Nicolas, Ferrero,
  Martens, and Barrat}}]{nicolas2018deformation}
\bibinfo{author}{\bibfnamefont{A.}~\bibnamefont{Nicolas}},
  \bibinfo{author}{\bibfnamefont{E.~E.} \bibnamefont{Ferrero}},
  \bibinfo{author}{\bibfnamefont{K.}~\bibnamefont{Martens}}, \bibnamefont{and}
  \bibinfo{author}{\bibfnamefont{J.-L.} \bibnamefont{Barrat}},
  \bibinfo{journal}{Reviews of Modern Physics} \textbf{\bibinfo{volume}{90}},
  \bibinfo{pages}{045006} (\bibinfo{year}{2018}).

\bibitem[{\citenamefont{Priesemann et~al.}(2014)\citenamefont{Priesemann,
  Wibral, Valderrama, Pr{\"o}pper, Le~Van~Quyen, Geisel, Triesch, Nikoli{\'c},
  and Munk}}]{priesemann2014spike}
\bibinfo{author}{\bibfnamefont{V.}~\bibnamefont{Priesemann}},
  \bibinfo{author}{\bibfnamefont{M.}~\bibnamefont{Wibral}},
  \bibinfo{author}{\bibfnamefont{M.}~\bibnamefont{Valderrama}},
  \bibinfo{author}{\bibfnamefont{R.}~\bibnamefont{Pr{\"o}pper}},
  \bibinfo{author}{\bibfnamefont{M.}~\bibnamefont{Le~Van~Quyen}},
  \bibinfo{author}{\bibfnamefont{T.}~\bibnamefont{Geisel}},
  \bibinfo{author}{\bibfnamefont{J.}~\bibnamefont{Triesch}},
  \bibinfo{author}{\bibfnamefont{D.}~\bibnamefont{Nikoli{\'c}}},
  \bibnamefont{and} \bibinfo{author}{\bibfnamefont{M.~H.} \bibnamefont{Munk}},
  \bibinfo{journal}{Frontiers in systems neuroscience}
  \textbf{\bibinfo{volume}{8}}, \bibinfo{pages}{108} (\bibinfo{year}{2014}).

\bibitem[{\citenamefont{Bauke}(2007)}]{bauke2007parameter}
\bibinfo{author}{\bibfnamefont{H.}~\bibnamefont{Bauke}}, \bibinfo{journal}{The
  European Physical Journal B} \textbf{\bibinfo{volume}{58}},
  \bibinfo{pages}{167} (\bibinfo{year}{2007}).

\end{thebibliography}

\end{document}